\documentclass[a4paper,12pt]{article}

\catcode`@=11
\def\secteqno{\@addtoreset{equation}{section}%
\def\theequation{\thesection.\arabic{equation}}}
\def\greek2{I\hspace{-.1em}I}
\def\beeqno{\begin{eqnarray*}}
\def\eneqno{\end{eqnarray*}}
\def\beeq{\begin{eqnarray}}
\def\eneq{\end{eqnarray}}
\def\*{e^{i\frac{\theta^{\mu\nu}}{2}\partial_\mu^\alpha \partial_\nu^\beta}}
\def\fra12{\frac{1}{2}}

\def\hsp1{\hspace{1cm}}

\def\={& \hspace{-.3cm}= \hspace{-.3cm}&}

\def\calh{{\cal H}}
\def\a{\alpha}
\def\b{\beta}
\def\d{\delta}
\def\g{\gamma}

\def\m{\mu}
\def\n{\nu}

\def\r{\rho}
\def\s{\sigma}

\def\e{\eta}
\def\p{\pi}
\def\P{\Pi}
\def\x{\xi}

\def\part{\partial}

\def\hep{hep-th/}
\def\nucl{Nucl. Phys. }
\def\phys{Phys. Rev. }

\def\modlet{Mod. Phys. Lett. }

\catcode`@=12
\secteqno

\begin{document}

\begin{titlepage}

\vspace{3cm}
\begin{flushright}
CHIBA-EP-133
\end{flushright}

\begin{center}
{\huge
Poisson Brackets, Strings  and Membranes}\\

\vspace{1cm}

Ken-Ichi Tezuka\footnote{e-mail: tezuka@physics.s.chiba-u.ac.jp}  \\

\vspace{1cm}

Graduate School of Science and Technology Chiba University, Japan \\

\vspace{5cm}

{\large  Abstract}   \\
\end{center}
We construct Poisson brackets at boundaries of open strings and membranes with 
constant background fields which are compatible with their boundary conditions. The 
boundary conditions are treated as primary constraints which give infinitely many 
secondary constraints. We show explicitly that we need only two (the primary one and one 
of the secondary ones) constraints to determine Poisson brackets of strings. We apply 
this to membranes by using canonical transformations.

\end{titlepage}
\section{Introduction}
Recently non-commutative spacetime attracted much attention from both theoretical 
\cite{seiberg}-\cite{ishibashirelation} and phenomenological \cite{tamaki} points of view. 
Especially in string theory, there are many studies of non-commutative 
descriptions of D-branes which are translated into a commutative description  by 
the Seiberg-Witten map \cite{seiberg}. 
\par
It was pointed out by Connes, Douglas 
and Schwarz \cite{connes} that M-theory with a constant background 3-form tensor 
field compactified on a torus can be identified with matrix theory compactified on 
a non-commutative torus. Similarly, string theory with a background NS 
B field is equivalent to string theory on a non-commutative space \cite{douglas}. 
\par
The non-commutativity comes from the fact that the 
canonical Poisson bracket does not coincide with a boundary condition 
\cite{ardalan}. Some 
authors have made efforts to obtain  Poisson brackets which are compatible with 
boundary conditions of strings \cite{ardalandirac}-\cite{chusuper} and 
membranes \cite{kawamoto,das}. Let 
us call them  a ``boundary Poisson brackets'' \footnote{After the first version of the 
present paper is finished, 
it is informed by Bering that the term ``boundary Poisson bracket'' has already 
been used in \cite{Solovev:1997qx}. Their definition of the term is 
different from ours.}. 
For strings, boundary Poisson brackets 
can be obtained by using the Dirac formalism 
\cite{ardalandirac,chu,kim}. The quantization is defined by replacement of a Dirac 
bracket with a commutator; $\{ \ , \ \}_D \to -i[ \ , \ ]$. When a NS B-field is 
turned on, $\{X^{\m},X^{\n}\}_D$ has non-zero value at boundaries, and boundary 
coordinates become non-commutative at the quantum level.    
\par
In M theory, the fundamental object is called  the ``M2-brane'' which is a 
2-dimensionally 
extended object. This is coupled with a 3-form field. A boundary condition of the membrane 
is non-linear, and it is difficult 
to get all of secondary constraints. Thus we need some  
ideas to deal with the system. In \cite{das}, a partial gauge fixing condition with  
which a boundary constraint of a membrane gives a finite number of constraints is 
introduced. We would like to determine a boundary Poisson bracket in a completely 
gauge fixed action. 
In this paper, keeping this in mind, we construct a boundary Poisson bracket of an 
open string by avoiding using all of secondary constraints, and apply this to a membrane. 
\par
The present paper is organized as follows. In the next section, we show that it is 
possible to  
construct a boundary Poisson bracket of an open string from two constraints by 
demanding that the canonical Poisson bracket is changed only at boundaries of an  
open string. In the Sect.3, we put to use the previous procedure in a system of 
a membrane. The section 4 is devoted to a summary. \\
\section{Strings and Constant 2-Form Fields}
In general, the canonical Poisson bracket;
\beeq
\{X^{\m}(\s),X^{\n}(\s')\}_p=0,\hspace{5mm}
\{X^{\m}(\s),P_{\n}(\s')\}_p=\d^{\m}_{\n}\d(\s-\s'), \hspace{5mm}
\{P_{\m}(\s),P_{\n}(\s')\}_p=0
\eneq 
in field theory and string theory with boundaries
does not coincide with their boundary conditions. In string theory, since open 
strings have endpoints, we have to impose their boundary conditions. Both Neumann and 
Dirichlet conditions change the canonical Poisson bracket at boundaries of open 
strings. In existence of a NS B-field, along a D-brane,  
a boundary condition of an open string becomes  a mixed type of Neumann and Dirichlet 
conditions. Also the mixed type boundary condition does 
not coincide with the canonical Poisson structure\cite{ardalan}. With respect to 
 mixed type boundary conditions , it is non-trivial work to determine  boundary 
Poisson brackets.
The representative method to construct boundary Poisson brackets is Dirac formalism.
\par
In the Dirac formalism, boundary conditions are dealt with as primary constraints. By 
the definition of a Dirac bracket, a boundary 
condition and a Dirac bracket are manifestly compatible with each other. In other 
words, a Dirac bracket between a boundary condition and canonical variables vanish, 
if the constraints are of second class. 
This system is one of the rare examples in which primary constraints produce 
infinitely many secondary constraints. 
\par
On this subject there are some papers \cite{ardalandirac,chu,kim}. 
In this system, we would like to see that we are able to determine a boundary Poisson 
bracket if we demand locality of the boundary Poisson bracket, i.e.~the  canonical 
Poisson bracket is changed only at the boundaries of an open string. In the following, we 
avoid using the Dirac formalism directly. 
\par
The gauge fixed action of a bosonic open string with a constant NS B field background is given 
by
\beeq
S=-\frac{T_s}{2}\int d^2\x [\part_{\a}X^{\m}\part^{\a}X_{\m}
-B_{\m\n}\epsilon^{\a\b}\part_{\a}X^{\m}\part_{\b}X^{\n}]
\eneq
where $T_s$ is the string tension. The canonical momentum is
$P^{\m}=T_s(\part_{\tau}X^{\m}+B^{\m \n}\part_{\s}X_{\n})$
and the action is non-singular. By the variational principle, the equation of motion 
is 
\beeq
\part_{\a}\part^{\a}X^{\m} =0   \label{5}
\eneq 
and along the D-brane the boundary condition
in terms of the canonical momentum  is 
\beeq
\int_0 ^{\p} d\s \d(\s)( \frac{1}{T_s}B^{\m \n}P_{\n}(\s)
+M^{\m \n}\part_{\s}X_{\n}(\s))=0.                 \label{010726-2}
\eneq
with $M=\e-B^2$ ($\e^{\m\n}$ is the target space flat metric tensor) and similarly  
with $\d(\s-\pi)$. The canonical Hamiltonian is 
\beeq
\calh=\frac{T_s}{2}[(\frac{1}{T_s}P_{\m}-B_{\m \n}\part_{\s}X^{\n} )^2
+\part_{\s}X^{\m}\part_{\s}X_{\m}].
\eneq
In this paper we will mainly consider the 
boundary condition only at $\s=0$ since the discussion of the condition at $\s=0$ is 
parallel with that at $\s=\pi$. Boundary Poisson brackets must be compatible with the 
constraint. We denote the boundary Poisson bracket as $\{ \ ,\ \}_b$ which is defined 
by the condition that the boundary Poisson bracket of the boundary condition with 
an arbitrary canonical variable vanishes;  
\beeq
\{f(X,P),\int^{\p}_{0} d\s' \d(\s')( \frac{1}{T_s}B^{\n \r}P_{\r}(\s')
+M^{\n \r}\part_{\s'}X_{\r}(\s'))\}_b=0       \label{1}   
\eneq
where $f(X,P)$ is an arbitrary function on the phase space.
From this condition only, we cannot determine the boundary Poisson bracket uniquely 
since there are only 2 independent equations for 3 unknowns \cite{ardalan}. In order to 
determine this uniquely, 
it was considered we must use the secondary constraints \cite{ardalandirac,chu}. 
The boundary constraint (\ref{010726-2}) gives infinitely many secondary 
constraints \cite{chu} which are
\beeq
&&\int_0 ^{\p} d\s \d(\s) \part_{\s}^{2n}( \frac{1}{T_s}B^{\m \n}P_{\n}(\s)
+M^{\m \n}\part_{\s}X_{\n}(\s))=0  \hspace{2cm}  (n=1,2,\cdots)    \label{15} \\
&&\hspace{3.9cm}\int _0 ^{\p} d\s \d(\s)\part_{\s}^{2n+1}P^{\m}(\s)=0 \hspace{2cm}  
(n=0,1,\cdots)  \label{16}.
\eneq
They originate from the condition of stationarity of the boundary constraint 
(\ref{010726-2}). From these we choose, for example, the $n=0$ case of (\ref{16}).  
We have another explanation for necessity of the condition for the case 
with $B_{\m\n}\neq 0$. We need the condition in order for the equation of motion 
(\ref{5}) to be equivalent to Hamilton's equations; 
\beeq
\part_{\tau}X^{\m}(\s)
&=&\frac{1}{T_s}P^{\m}(\s)-B^{\m \n}\part_{\s}X_{\n}(\s)            \label{6} \\
\part_{\tau}P^{\m}(\s) 
&=&B^{\m \n}\part_{\s}P_{\n}+T_s M^{\m\n}\part^2_{\s}X_{\n}.
                            \label{9}
\eneq
At the boundaries, the equation (\ref{6}) can also be rewritten as 
\beeq
\left. \part_{\tau}X^{\m}(\s)\right|_{\s=0,\p} 
=  \frac{1}{T_s}\left.(M^{-1})^{\m \n}P_{\n}(\s) \right|_{\s=0,\p}    \label{7}
\eneq
By virtue of the condition (\ref{16}) with $n=0$, the equation of motion (\ref{5}) is 
equivalent to the equations (\ref{7}) and (\ref{9}). 
\par     
Here we have a question. Do we need the all of the secondary constraints to construct 
the boundary Poisson bracket? In \cite{chunoncom} only the equation of motion (\ref{5}) 
and the boundary condition (\ref{010726-2}) are 
used. We would like to consider this in the following.
\par
Let us add the condition  
\beeq
\{f(X,P),\int_0^{\p} d\s' \d(\s')\part_{\s'}P^{\n}(\s') \}_b =0   \label{11} 
\eneq
to the previous one (\ref{1}). 
Since we have 4 equations for 3 unknowns, in spite of existence of the 
infinitely many constraints, we have a possibility to be able to determine the boundary 
Poisson bracket, if we demand their locality. 
We assume that the bracket is anti-symmetric and bilinear, and satisfies the 
derivation rule which are fundamental properties of the canonical Poisson bracket.    
\par
By solving (\ref{1}), (\ref{11}) with $f=X,P$ we have 
\beeq
\{X^{\m}(\s),X^{\n}(\s')\}_b&=&\frac{1}{T_s}Q(\s,\s')(M^{-1}B)^{\m\n}  \label{3}\\
\{X^{\m}(\s),P_{\n}(\s')\}_b&=&\d^{\m}_{\n}\hat{\d}(\s,\s') \\
\{P_{\m}(\s),P_{\n}(\s')\}_b&=&0   \label{4}
\eneq
where 
\beeq
\left.\part_{\s}\hat{\d}(\s,\s')\right|_{\s=0,\p}=0,  \hspace{2cm}
\part_{\s}Q(\s,\s')=\hat{\d}(\s,\s').    \label{2-2}
\eneq
Since $\hat{\d}(\s,\s')$ have to be equivalent to the ordinary delta function  
$\d(\s-\s')$ at bulk, we use the ansatz;
\beeq
\hat{\d}(\s,\s')=\d(\s-\s')+a_1\d(\s+\s')+a_2\d(\s+\s'-2\p)
\eneq
where $a_1$ and $a_2$ are constants to be determined. In the neighborhood of $\s=0$, 
\beeq
\hat{\d}(\s,\s')=\d(\s-\s')+a_1\d(\s+\s').
\eneq 
Although the delta function is not a periodic function, we calculate the Fourier 
expansion as if this has periodicity $2\p$. The Fourier expansion of the 
delta functions are 
\beeq
\d(\s-\s')&=&\frac{1}{\pi}
     [1+2\sum_{n =1}^{\infty}(\cos n\s\cos n\s'+\sin n\s\sin n\s')]   \\
\d(\s+\s')&=&\frac{1}{\pi}
      [1+2\sum_{n =1}^{\infty}(\cos n\s\cos n\s'-\sin n\s\sin n\s')]  
\eneq
In order to satisfy the left equation of (\ref{2-2}), $a_1=1$. Similarly we have $a_2=1$. 
Then we have
\beeq
\hat{\d}(\s,\s')=\d(\s-\s')+ \d(\s+\s')+\d(\s+\s'-2\p).
\eneq
From this and  (\ref{2-2}) we obtain
\beeq
Q(0,0)=-Q(\p,\p)=-2
\eneq
where we have assumed $Q(0,\p)$ and $Q(\p,0)$ are zero.
\par
It is easy to find that the boundary Poisson bracket and other constraints  
(\ref{15}) and (\ref{16}) are consistent. 
The procedure will be applicable for other systems with boundary conditions which are 
linear in canonical variables.     
\section{Membranes and Constant 3-Form Fields}
As a next step we consider a boundary Poisson bracket of an open 
membrane. Since membranes are 2-dimensionally extended objects, they are coupled with 
3-form fields. Our aim in the present section is to construct a boundary Poisson 
bracket for 
an open membrane with a constant 3-form C-field background. Due to the 3-form field, 
a boundary term appears in the action of an open membrane, which is third order in its 
membrane's coordinates. So its boundary condition becomes non-linear and becomes a mixed 
type condition. By the non-linearity, a conventional Dirac procedure is not 
easy task\cite{kawamoto,das}. It is hard to construct all of secondary 
constraints.  We follow the 
previous procedure also in the present case. In addition to this, we use
canonical transformations in order to make calculations possible.
\par
We consider a membrane whose topology is cylindrical. The worldvolume coordinates of 
the membrane are parameterized by $\tau$, $\s_1\in[0,\p]$ and $\s_2 \in [0,2\p]$. 
Along the $\s_2$ direction the membrane is periodic.  
\par
We use the Polyakov action \cite{taylor};
\beeq
S=-\frac{T_m}{2} \int d^3 \x [
\sqrt{-\det \g_{\a\b}}(\g^{\a\b}\part_{\a}X^{\m}\part_{\b}X_{\m}-1)
+\frac{2}{3!}\epsilon^{\a\b\g}C_{\mu \nu \rho}\part_{\a}{X^{\mu}}\part_{\b}{X^{\nu}}
\part_{\g}{X^{\rho}}]. \label{010726-15}
\eneq
where $d^3 \x \equiv d\tau d\s_1d\s_2$, $T_m$ is the membrane tension and 
$\g_{\a\b}$ is the metric tensor on the membrane. $C_{\m\n\r}$ is a constant background 
field. This action has the worldvolume reparametrization 
invariance. Then we need to perform a gauge fixing procedure. Let us adopt the gauge 
condition; $\g_{0a}=0\  \g_{00}=-\det h_{ab}$ with $a,b=1,2$. Here $h_{ab}$ is the 
induced metric on the membrane. The gauge fixed action is 
\beeq
S=\frac{T_m}{2}\int d^3 \x [\part_{\tau}{X^{\mu}}\part_{\tau}{X_{\mu}}-\fra12
\{X^{\mu},X^{\nu}\}\{X_{\mu},X_{\nu}\}
-\frac{2}{3!}\epsilon^{\a\b\g}C_{\mu \nu \rho}\part_{\a}{X^{\mu}}\part_{\b}{X^{\nu}}
\part_{\g}{X^{\rho}}]          \label{12}
\eneq
where $\{f,g\}=\epsilon^{ab}\part_{a}f\part_{b}g$. 
\par
The canonical Poisson bracket for the cylindrical membrane is changed by a boundary 
condition, 
and also by the fact that the membrane is periodic along the $\s_2$ direction.    
Before studying the boundary condition at $\s_1=0,\p$, we see the modification of the 
canonical Poisson bracket due to the periodicity along the $\s_2$ direction. 
\par
The left hand side of the canonical Poisson bracket;
\beeq  
\{X^{\m}(\x),P_{\n}(\x')\}_p=\d^{\m}_{\n}\d(\x-\x')
\eneq
has periodicity $2\p$ along the $\s_2$ and $\s_2'$. However, the right hand side does 
not have such periodicity. So the delta function $\d(\s_2-\s_2')$ have to be replaced 
by a periodic one $\tilde{\d}(\s_2-\s_2')$ which satisfies
\beeq
\tilde{\d}(\s_2-\s_2')=\tilde{\d}(\s_2-\s_2'+2\p m) \hsp1 (m \in \mathbf{Z}).
\eneq
Note that the canonical Poisson bracket for the cylindrical membrane is not changed only 
at the boundaries of the membrane but also at the bulk of it.       
\par
Next we would like to see a change of the canonical Poisson bracket by a boundary 
condition. 
For the action (\ref{12}), the canonical Hamiltonian is 
\beeq
\calh=\frac{T_m}{2}(\frac{1}{T_m}\P_{\m}+\fra12 C_{\m \n \r}\{X^{\nu},X^{\r}\})^2
+\frac{T_m}{4}\{X^{\m},X^{\n}\} ^2.   \label{3-1}
\eneq
Notice that the Hamilton theory (\ref{3-1}) is canonically equivalent to that with  
$C_{\m\n\r}=0$. They are related by the transformation
\beeq
\tilde{\P}_{\m} = {\P}_{\m}+\fra12 C_{\m \n \r}\{X^{\nu},X^{\r}\}, \hspace{1cm}   
\tilde{X}^{\m} = X^{\m}. 
\eneq
We will start from the system $(\tilde{X},\tilde{\P})$ with $C_{\m\n\r}=0$.
\par
With respect to the membrane, we adopt the boundary condition 
$\part_{\s_1}\tilde{X}^{\m}|_{\s_1=0,\p}=0$. The secondary constraint is  
$\part_{\s_1}\tilde{\P}^{\m}|_{\s_1=0,\p}=0$. Furthermore the phase space $(X,\P)$ can be  
transformed by $X^{\m} = \hat{\P}^{\m}$ and $\P^{\m} = -\hat{X}^{\m}$ which are 
canonical ones. In terms of the canonical variables $(\hat{X},\hat{\P})$, the conditions 
are replaced by
\beeq
\left. \part_1 \hat{\P}^{\m}\right|_{\s_1=0,\p} \= 0    \label{3-2}         \\
\left. \part_1\hat{X}_{\m}-C_{\m \n \r}\part_1 ^2\hat{\P}^{\nu}\part_2\hat{\P}^{\r} 
\right|_{\s_1=0,\p}\= 0  .     \label{3-3}
\eneq
In order to determine a boundary Poisson bracket of the membrane, we follow the method 
used in the previous section with the 
conditions (\ref{3-2}) (\ref{3-3}). The result is   
\beeq
\{\hat{\P}_{\m}(\x),\hat{\P}_{\n}(\x')\}_{b}\=0   \\
\{\hat{X}^{\m}(\x),\hat{\P}_{\n}(\x')\}_{b}\=\d^{\m}_{\n}\hat{\d}(\x-\x')  \\
\{\part_1 \hat{X}_{\m}(\x),\hat{X}_{\n}(\x')\}_b\=-T_mC_{\m\n\r}
[\part_1^2\hat{\d}(\x-\x')\part_2\hat{\P}^{\r}(\x)-\part_1 ^2\hat{\P}^{\r}
\part_2\hat{\d}(\x-\x')]
\eneq 
with $\hat{\d}(\x-\x'):=\hat{\d}(\s_1,\s_1')\tilde{\d}(\s_2-\s_2')$. The last one is 
valid only at $\s_1=0,\p$. We have obtained the boundary Poisson bracket 
of the membrane without using any approximations. When the $C$-field has non-zero value, 
the boundaries of the membrane becomes non-commutative.   
\section{Summary}
In this paper, we have explicitly shown that though there exists infinitely many 
secondary constraints, the boundary Poisson bracket of a bosonic open string can 
be determined only from two constraints by demanding its locality. 
The brackets (\ref{3})-(\ref{4}) coincide with all other secondary 
constraints (\ref{15}) and (\ref{16}). In other words, the boundary Poisson bracket 
between canonical variables and the secondary constraints vanish. 
\par
Secondly we have applied the procedure
to the system of a membrane with a constant 3-form field background 
by using the canonical transformations. The non-commutativity depends on the canonical 
momentum, and it is first order in $C_{\m\n\r}$.  

\vspace{1cm}

The author would like to thank Prof.\hspace{-.5em} Tadahiko Kimura for helpful discussion and 
careful reading of manuscript.

\end{document}